\documentclass[letter,fleqn,usenatbib]{mnras}

\usepackage{newtxtext,newtxmath}
\usepackage[T1]{fontenc}
\usepackage{ae,aecompl}

\usepackage{rotating}
\usepackage{graphicx}	% Including figure files
\usepackage{amsmath}	% Advanced maths commands
\usepackage{amssymb}	% Extra maths symbols

\usepackage[switch]{lineno}
\usepackage{graphicx}
\usepackage{multirow}
\usepackage{ amssymb }
\usepackage{soul}
\usepackage{xspace}
\usepackage{pdflscape}

\newcommand{\Rsun}{\mbox{$\rm R_{\odot}$}}

\newcommand{\num} {$\nu_{\rm max}$\xspace}

\newcommand{\Kepler} {\textit{Kepler}\xspace}
\newcommand{\kepler} {\textit{Kepler}\xspace}

\newcommand{\KIC}[1]{{KIC\,#1\xspace}}

\newcommand{\teff}{$T_{\mathrm{eff}}$\xspace}

\newcommand{\varR}{$\varepsilon_r$\xspace}
\newcommand{\varr}{$\varepsilon_r$\xspace}
\newcommand{\varC}{$\varepsilon_{\rm crit}$\xspace}

\newcommand{\ddm}{\mathrm{d}}

\usepackage[switch]{lineno}
\newcommand{\new}[1]{\textbf{\cya{\bf#1}}}

\newcommand{\linguistic}[1]{\textit{\cya{#1}}}

\usepackage{xcolor}
\definecolor{CornBlue}{RGB}{100,149,237} %upper case 'RGB' takes values > 1
\definecolor{SteelBlue}{RGB}{65,105,225}
\newcommand{\steelBlue}{\textcolor{SteelBlue} }

\renewcommand{\new}[1]{{#1}}
\renewcommand{\new}[1]{{\steelBlue{#1}}}
\renewcommand{\linguistic}[1]{\textit{\steelBlue{#1}}}

\newcommand{\Figure}[1]{Fig.\,\ref{#1}\xspace}
\newcommand{\Table}[1]{Table\,\ref{#1}\xspace}

\newcommand{\Section}[1]{Section\,\ref{#1}\xspace}

\title[Circularization of giant stars in the \Kepler sample]{Testing tidal theory for evolved stars by using red-giant binaries observed by \textit{Kepler}}

\author[P. G. Beck et al.]{
P.\,G.~Beck,$^{1,2}$ %\thanks{paul.beck@iac.es}
S.~Mathis,$^{3,4,5}$
F. Gallet,$^{6}$ C. Charbonnel,$^{6,7}$ 
\newauthor
~M. Benbakoura,$^{3,4}$ 
R.\,A. Garc{\'i}a$^{3,4}$
\& J.-D. do Nascimento, Jr.$^{8,9}$\\
~$^{1}$ Instituto de Astrof\'{\i}sica de Canarias, E-38200 La Laguna, Tenerife, Spain\\%1
~~$^{2}$ Departamento de Astrof\'{\i}sica, Universidad de La Laguna, E-38206 La Laguna, Tenerife, Spain\\%2
~~$^{3}$ {IRFU, CEA, Universit\'e Paris-Saclay, F-91191 Gif-sur-Yvette, France}\\%3
~~$^{4}$ {Universit\'e Paris Diderot, AIM, Sorbonne Paris Cit\'e, CEA, CNRS, F-91191 Gif-sur-Yvette, France}\\%4
~~$^{5}$ LESIA, Observatoire de Paris, PSL Research Univ., CNRS, Univ.\,Pierre et Marie Curie, Univ.\,Paris Diderot,  92195\,Meudon, France\\%5
~~$^{6}$ Department of Astronomy, University of Geneva, Chemin des Maillettes 51, 1290, Versoix, Switzerland\\%6
~~$^{7}$ IRAP, UMR 5277, CNRS and Universit{\'e} de Toulouse, 14 Av. E. Belin, 31400, Toulouse, France\\%7
~~$^8$ Harvard-Smithsonian Center for Astrophysics, Cambridge, MA 02138, USA\\%8
~~$^9$ Departamento de F\'isica, Universidade Federal do Rio Grande do Norte, CEP: 59072-970 Natal, RN, Brazil%9
}
\date{submitted: 24 December 2017, accepted: 18 June 2018}%Accepted XXX. Received YYY; in original form ZZZ}

\pubyear{2018}

\begin{document}
\label{firstpage}
\pagerange{\pageref{firstpage}--\pageref{lastpage}}
\maketitle

\begin{abstract}
Tidal interaction governs the redistribution of angular momentum in close binary stars and planetary systems and determines the systems evolution towards the possible equilibrium state. Turbulent friction acting on the equilibrium tide in the convective envelope of low-mass stars is known to have a strong impact on this exchange of angular momentum in binaries. Moreover, theoretical modelling in recent literature as well as presented in this paper suggests that the dissipation of the dynamical tide, constituted of tidal inertial waves propagating in the convective envelope, is weak compared to the dissipation of the equilibrium tide during the red-giant phase. This prediction is confirmed when we apply the equilibrium-tide formalism developed by Zahn (1977), Verbunt\,\&\,Phinney (1995), and Remus,\,Mathis\,\&\,Zahn (2012) onto the sample of all known red-giant binaries observed by the NASA \Kepler mission. Moreover, the observations are adequately explained by only invoking  the equilibrium tide dissipation. Such ensemble analysis also benefits from the seismic characterisation of the oscillating components and surface rotation rates.  Through asteroseismology, previous claims of the eccentricity as an evolutionary state diagnostic are discarded. This result is important for our understanding of the evolution of multiple star and planetary systems \hbox{during advanced stages of stellar evolution.}
\end{abstract}

%%%%%%%%%%%%%%%%%%%%%%%%%%%%%%%%%%%%%%%%%%%%%%%%%%

\section{Introduction}
\vspace{-2mm}
A binary system has reached the state of minimum energy through tidal interactions when the orbit is circularised and the orbital and stellar spins are aligned and synchronised 
\citep[e.g.][]{Hut1980,Zahn2013}. 
While for orbital periods shorter than $\sim$10 days nearly all systems are circularised, a large spread in eccentricity is found for longer orbital periods \citep[e.g.][]{Mazeh2008}. Consequently, for wider binaries the orbital period alone is not sufficient to characterise tidal interactions 
\citep[hereafter VP95; \citealt{Torres2010AccurateMasses}]{Verbunt1995}. 
The efficiency of the physical processes that dissipate the kinetic and potential energies of tides into heat 
depends on stellar structure (in particular, the respective size and mass of the radiative and convection zones), fundamental parameters, and rotation. 
This determines the evolutionary pace for an eccentric binary towards equilibrium \citep[e.g.][]{Zahn1977, OgilvieLin2007, Mathis2015, Gallet17}. 

In stars, tides can be understood as the combination of an equilibrium/non wave-like tide and of a dynamical/wave-like tide \citep[e.g.][]{Zahn1977, Ogilvie2013, Ogilvie2014}. The equilibrium tide is a large-scale flow driven by the hydrostatic adjustment of a star induced by the tidal gravitational perturbation due to the companion \citep[e.g.][]{zahn1966, remus2012, Ogilvie2013}. It is efficiently dissipated in the convective envelope of low-mass stars during their evolution \citep[e.g.][]{Verbunt1995,ZB1989} because of the friction applied on its velocity field by turbulent convection \citep[e.g.][]{zahn1966,Mathis16}. However, the equilibrium tide is not a solution of the complete equations of stellar hydrodynamics and it must be completed by the so-called dynamical tide introduced by \cite{Zahn1975}. In rotating stars, it is constituted by tidal inertial waves that propagate in convective regions and are driven by the Coriolis acceleration \citep[e.g.][]{OgilvieLin2007,Ogilvie2013,Mathis2015} and by tidal gravito-inertial waves that are excited in stably stratified radiative zones and driven by buoyancy and the Coriolis acceleration \citep[e.g.][]{Zahn1975,Terquem98,OgilvieLin2007,Barker10}. They are damped by the convective turbulent friction and thermal diffusion and breaking, respectively.

\begin{figure}
\centering
\includegraphics[width=1\columnwidth]{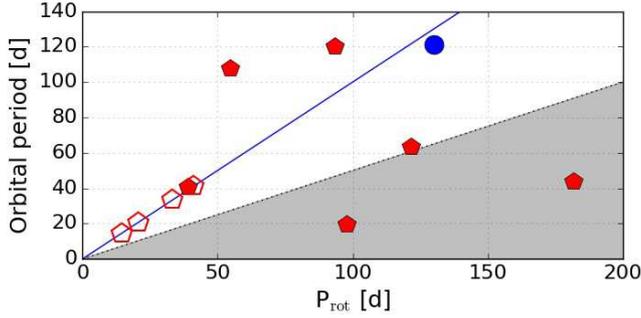}
\caption{Orbital versus surface rotation period of \Kepler red-giant binaries. 
The \linguistic{grey} shaded area depicts the prohibited region for the dynamical tide, i.e., $P_{\rm orb}<\frac{1}{2}P_{\rm rot}$. 
Red pentagons and blue dots mark eclipsing (G14,\,G16) and heartbeat systems (B14,\,B18) observed by \Kepler, respectively. Open symbols denote non-oscillating primaries.
\label{fig:inertialWaves}}
\end{figure}

Because of the long tidal evolution time scales compared to the existing time bases of observational data, such theoretical descriptions of tidal interactions within binary stars are difficult to probe empirically. VP95 however tested successfully their formalism for the dissipation 
of the equilibrium tide in evolving convective envelopes using a sample of spectroscopic red-giant binary systems located in open clusters. 
Today, this test can be extended, thanks to the NASA \Kepler space mission, originally designed for detecting planetary transits \citep{Borucki2010} which has been very successful in finding binary systems.
Among the 197\,100 stars observed by \Kepler \citep{Mathur2017}, $\sim$3000 binary stars are known \citep[][]{Kirk2016}. Thirty-five of these systems contain a red-giant component. Through combination of space photometry, asteroseismology, and ground-based spectroscopy, these systems are 
well characterised 
(\citealt{Frandsen2013}, \citealt{Gaulme2014, Gaulme2016},
\citealt{Rawls2016},
\citealt{Beck2014a,Beck2017b}, hereafter F13,\,G14,\,G16,\,R16,\,B14,\,\&\,B18). The main catalogs were presented by G14\,\&\,G16 and B14 (Tab.\,\ref{tab:LiteratureParameters}). While G14\,\&\,G16 focused on eclipsing systems, B14 presented binary stars exhibiting photometric signatures of tidal interactions due to the non-adiabatic component of the equilibrium tide, as predicted theoretically by \cite{Kumar1995}. We refer to them as heartbeat stars, a term coined by \cite{Thompson2012}.

In this work, we constrain the relative strength of the dissipation of the equilibrium and dynamical tides in red giant stars. We use the formalisms developed by VP95 and \cite{remus2012} for the equilibrium tide as well as \cite{Ogilvie2013} and \cite{Mathis2015} for the dynamical tide and confront their predictions to the observed orbital properties of the  \Kepler red-giant binaries sample. We discuss tidal interactions in these systems in the light of surface rotation and seismology provided~by~\Kepler.
\vspace{-2mm}

\section{Tidal dissipation in red giant stars} \label{sec:Description}

\subsection{Equilibrium tide \label{sec:equilibriumTide}}

For stars with large convective envelopes such as red giant stars, the first of the main tidal dissipation mechanisms is the turbulent friction applied by convection on the equilibrium tide \citep[e.g.][]{zahn1966,remus2012}. 

In the Appendix\,\ref{AppendixB}, we provide the expression of this dissipation ${\mathcal D}_{\rm eq}$ as a function of the stellar structural properties and rotation and of the orbital period. It will be used to compare the respective strength of the equilibrium and dynamical tides on the RGB. In addition, we follow 
VP95  who rewrote the associated circularisation timescale 
$\tau_{\rm circ}$ derived by \cite{Zahn1977} in the following way:
\begin{eqnarray}
\label{eq:dedt}
\frac{1}{\tau_{\rm \new{circ}}} = \frac{d \ln e}{dt}&\simeq&
-1.7 f 
\cdot \left(\frac{T_{\rm eff}}{4500\,K}\right)^{4/3} 
\cdot \left(\frac{M_{\rm env}}{M_\odot}\right)^{2/3} \\
&&~\cdot~\frac{M_\odot}{M_1}
\cdot \frac{M_2}{M_1}
\cdot \frac{M_1+M_2}{M_1}
\cdot \left(\frac{R_1}{a}\right)^8~~ [{\rm yr}^{-1}]\,, \nonumber
\end{eqnarray}
whereby the parameters of the primary and the secondary are the  mass $M$  and the radius $R$. The effective temperature and mass of the convective envelope of the giant component are given by $T_{\rm eff}$ and $M_{\rm env}$, respectively. The dimensionless parameter $f$ is of the order of unity and determined by the details of the  convective turbulent friction. Finally, $a$ is the system's semi-major axis. 
\begin{figure}
\centering
\includegraphics[width=1\columnwidth,height=45mm]{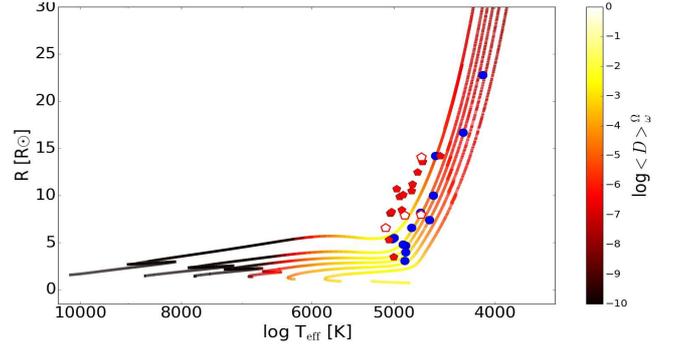}
\caption{Dissipation of the dynamical tide as a function of stellar evolution. The colour-code reflects the intensity of the dissipation along the theoretical tracks starting at the zero-age main sequence in the $R_{*}$-\teff plane, calculated for the sample mass range (Z$_{\odot}$ tracks for 0.8,\,1.0, 1.2,\,1.4,\,1.6,\,1.8,\,2.0\,\&\,2.2\,M$_{\odot}$).\ 
The same symbols as in Fig.\,\ref{fig:inertialWaves} are used.
\label{fig:RTeff}}
\end{figure}

To quantify the tidal dissipation within the giant component in detached  close binary systems and classify them,  VP95 isolate the terms of Eq. (\ref{eq:dedt}) that depend on the parameters of the primary component and introduce the circularisation function
\begin{equation}
{I\left(t\right) = \int_0^t\left(\frac{T_{\rm eff}}{4500\,K}\right)^{4/3} 
\cdot \left(\frac{M_{\rm env}(t')}{M_\odot}\right)^{2/3}
\cdot \left(\frac{R_1(t')}{R_\odot}\right)^{8}
dt'\,.}
%\nonumber
\label{eq:IofT}
\end{equation} 
By integrating Eq. (\ref{eq:IofT}) along the evolutionary tracks of representative stellar models, VP95 found that $I(t)$ is a monotonic function of the stellar radius  
that can be well approximated by the  analytical expression
\begin{equation}
I = 7.6\cdot10^8\cdot\left(\frac{R_1}{R_\odot}\right)^{6.51} [{\rm yr}]\,.
\label{eq:IofR}
\end{equation}

For stars in the red-giant phase, the radius is the parameter that varies most drastically with time, more than two orders of magnitude, while the mass  of the convective envelope and the effective temperature change only modestly.
By using Eq.\,(\ref{eq:dedt}), Eq.\,(\ref{eq:IofT}), and Kepler's third law, VP95 derive the rate of eccentricity reduction, %$\Delta \ln e$,  
\begin{eqnarray}
\label{eq:deltaLnE}
\frac{\Delta \ln e}{f}
 &=& \frac{-1.7}{ 10^{5}}\cdot\left(\frac{M_1}{M_\odot}\right)^{-11/3}\cdot 
 \frac{q}{(1+q)^{5/3}} \cdot I(t) \cdot \left(\frac{P_{\rm orb}}{{\rm day}}\right)^{-16/3}\,.
\end{eqnarray} 
with $q$=$M_2$/$M_1$  the components mass ratio.
In their plots, VP95 use the  logarithm of negative value of Eq.\,(\ref{eq:deltaLnE}), which, for the convenience of the reader, we call in this work $\varepsilon_{\rm r}$. 
\begin{eqnarray} 
\label{eq:varr}
\varepsilon_{\rm r} =\log \left[-\left(\frac{\Delta \ln e }{ f}\right)\right]\,.
\end{eqnarray}
This quantity represents the inverse time scale of tidal circularisation as a diagnostic of the strength of the dissipation. We will compute it for our sample in \Section{sec:Circularization}.

\begin{figure*}
\centering
\includegraphics[width=\textwidth,height=45mm]{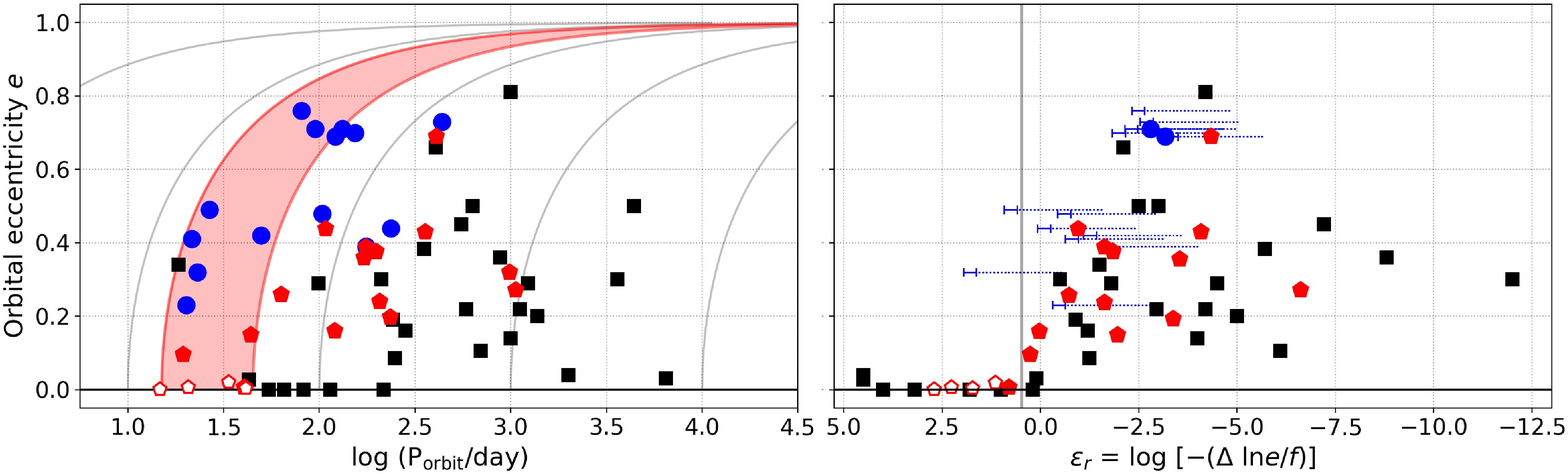}
\vspace{-4mm}
\caption{
Eccentricity of red-giant binaries in the \Kepler sample and open clusters. Red pentagons and blue dots mark \Kepler red giants in eclipsing (G14, G16) and heartbeat systems (B14, B18), respectively. Open symbols denote non-oscillating primaries. The sample of VP95 cluster stars is shown as black squares.
 \textit{Left:} $e$ versus orbital period. Grey lines indicate tracks of constant angular momentum for circularised systems with periods of 10, 10$^2$, 10$^3$, and 10$^4$ days.  The red shaded area marks the range of periods, 15$\leq$P$_{\rm circ}$$\leq$45 days. 
\textit{Right:} $e$ versus the circularisation function \varr. Following VP95, the horizontal axis is inverted.
Filled symbols mark systems with known mass ratio, while blue solid 'H'-symbols indicate the \varr-range, calculated for the mass ratio range, 1\,$\leq$\,$q$\,$\leq$\,1/5, and blue dotted lines represent a mass ratio between unity and q$\simeq$1:1000. The vertical grey line depicts the value of $\varepsilon_{\rm crit}$. 
\label{fig:eccentricities}}
\vspace{-2mm}
\end{figure*}

\vspace{-1mm}
\subsection{Dynamical tide}\label{sec:dynamicalTide}
\vspace{-1mm}
In the deep convective envelope of red giant stars, 
the dynamical tide is constituted by tidal inertial waves driven by the Coriolis acceleration
\footnote{As a first step, we do not take into account the tidal dissipation in the radiative stellar core.}
if $P_{\rm orb}$\,>\,$\frac{1}{2}P_{\rm rot}$, with \hbox{$P_{\rm orb}$ and $P_{\rm rot}$} the periods of the orbital motion and stellar surface rotation, respectively \citep{OgilvieLin2007}\footnote{In the case of eccentric systems, other tidal frequencies will be excited \citep[see Appendix\,\ref{AppendixA} and][]{Zahn1977,Ogilvie2014}.}. 
As shown in Fig.\,\ref{fig:inertialWaves}, most of the eleven \Kepler binary systems with known rotation periods (G14,\,G16,\,B14,\,B18, see Tab.\,\ref{tab:LiteratureParameters}) lie in the regime where tidal inertial waves can propagate. 
  
We compute the structural frequency-averaged dissipation \hbox{$<$$\mathcal{D}$$>_{\rm struct.}$}\,(see Eq.\,\ref{dissipequabis} and Eq.\,3 in \citealt{Gallet17}), introduced by \citet{Ogilvie2013} and \citet{Mathis2015}  
for standard (non-rotating) solar metallicity stellar models {in the mass range 0.8\,$\leq$\,M/M$_\odot$\,$\leq$\,2.2 (covered by the sample of \Kepler stars, see Tab.\,\ref{tab:LiteratureParameters} in the Appendix) 
computed with the stellar evolution code STAREVOL \citep[see][and references therein]{Gallet17}. 
As shown in \Figure{fig:RTeff}, the dissipation varies \linguistic{by} several orders of magnitude along the stellar evolution from the main-sequence to the Red Giant Branch (hereafter RGB) phase
 \citep[see][and  Appendix\,\ref{AppendixA}]{OgilvieLin2007}.

We see that for stars with solar-metallcity and masses above 1.4\,M$_\odot$ the structural frequency-averaged tidal dissipation is most efficient at the end of the sub-giant branch, while for stars with 0.8$\leq$M/M$_\odot$$\leq$1.4 it is efficient throughout  the main-sequence and sub-giant branch. However the efficiency is reduced by several orders of magnitude, once  
the stars passed the bottom of the RGB. This diagram only shows the efficiency of dissipation due to the dynamical tide. 
From seismic analysis and comparison with the evolutionary tracks in the HRD it is shown that all our sample stars are on the RGB (see\,Sec.\,\ref{sec:synergies}), with half of them lying in or just above the region of maximum dissipation. 

On the their way towards the RGB, stars experience a rapid expansion of their convective envelope in mass to maintain a surface hydrostatic equilibrium. This results in a drastic reduction of the efficiency of the dissipation of tidal inertial waves 
because the stellar structure configuration tends towards a full-sphere configuration. Here the viscous friction acting on sheared inertial waves attractors become less efficient because inertial waves become regular \citep[we refer to the detailed discussion in][]{Mathis2015}. Simultaneously, the growing (in radius) envelope amplifies the strength of the dissipation of the equilibrium tide as shown by Eq. (\ref{eq:dissip_end}).

\vspace{-2mm}
\section{Comparison with VP95 and \Kepler samples}\label{sec:Circularization}

VP95 applied their formalism to 28 well-studied binaries with giant components with  masses derived from cluster isochrones. In \Figure{fig:eccentricities}, the orbital eccentricity $e$ of the systems in the cluster-sample is plotted as a function of the orbital period and \varR as black squares.
While in the $e$-$P$ plane no convincing structure is found, plotting the eccentricity versus \varR  separates efficiently the eccentric from the circularised systems. 
From their analysis, VP95 suggest 
that a value of [$\Delta \ln e/f$]\,$\simeq$\,3 
marks the approximate transition limit between quasi- and non-circularised systems.  Following Eq.\,(\ref{eq:varr}), we refer to this value as $\varepsilon_{\rm crit}$\,$\simeq$\,0.478, which is marked as a vertical line in the right hand panel of Fig.\,\ref{fig:eccentricities}. 
We note that \varC is not indicating an exact dichotomy between circularised and eccentric systems, but can be understood as the entry line to the rapid circularisation phase. Figure\,\ref{fig:eccentricities} suggests that \varC is lower for very eccentric systems.

The red-giant binaries found by the \Kepler mission now provide a key comparison data set. The masses of the components are determined from seismology and dynamical approaches, complementing mass determination through isochrone fitting as done for the cluster stars. The literature parameters of the stars in the \Kepler sample are given in Tab. \ref{tab:LiteratureParameters}.
In the left panel of \Figure{fig:eccentricities}, eclipsing systems (F13, G14, G16) are depicted as red pentagons, while heartbeat stars (B14) are shown as blue dots.

From the full sample of \Kepler red-giant binary systems, a subset of 19 stars were found by various studies to be double-lined spectroscopic binary (SB2) systems (F13, G16, R16, B18), providing us with  precise mass ratios of the system. By comparing the seismic and the dynamical masses of the SB2, G16 showed that masses and radii derived from seismic scaling relations are overestimated by $\sim$15\% and 5\%, respectively. 
Therefore, we use, whenever available, the dynamical or corrected seismically inferred masses.
For 14 out of the 18 heartbeat stars, B14 reported orbital eccentricities, based on their spectroscopic radial velocity (RV) monitoring. 
For one heartbeat system, KIC\,5006817, the mass ratio was constrained from a combined fit of the \Kepler light and  RV curve (B14). With the exception of KIC\,913796 (B18), these systems have typically much fainter companions stars, making it  observationally challenging to detect the secondary's spectrum. 
For the twelve remaining heartbeat systems with known seismic masses but unknown mass ratio, we assumed a mass ratio 1/5\,$\leq$\,$q$\,$\leq$1 (values taken from B18 and B14). 
To explore the expected range of \varR for planetary systems, the mass ratio down to 1/1000 is depicted, representing an approximate mass ratio for a star hosting a giant planet. We note that in \Figure{fig:RTeff} systems from G16 are systematically hotter than those from B14. Such difference could overestimate the seismic masses and radii in G16 of a few percent but does not explain the full 15\%.

The \Kepler and the VP95-cluster samples are complementary in terms of primary mass and orbital periods, covered by each sample. With a mean primary mass of 2.0$\pm$0.4\,M$_\odot$ the stars in the VP95 sample are substantially more massive than stars in the \Kepler sample (1.3$\pm$0.3\,M$_\odot$), but are very similar in terms of the mass ratio, 0.6$\pm$0.3 and 0.7$\pm$0.3, respectively. As shown in \Figure{fig:eccentricities} (left panel), all except two 
systems of VP95 have orbital periods $P_{\rm orbit}$\,$\gtrsim$\,45\,d, and most eccentric systems are found at even periods of several hundred or thousand of days.  Such long periodic systems are hard to detect from the $\sim$1400\,d of 
\Kepler photometry.
The \Kepler sample however contains systems with a wide range of eccentricities (0\,$\leq$\,$e$\,$\lesssim$\,0.8) at $P_{\rm orbit}$\,$\leq$\,45\,d.
 Combining both samples allows us to explore tides in short and long period systems. 
It is interesting to note that both samples are lacking systems with 0.5\,$\lesssim$\,$e$\,$\lesssim$\,0.7.  
While the \Kepler and cluster samples are different in their distribution in period (\Figure{fig:eccentricities}, left panel), viewing both samples in the $e$-\varR plane leads to a very homogeneous pattern (right panel) with circularised systems (excepting KIC\,2720096 and KIC\,8095275) if $\varepsilon>\varepsilon_{\rm crit}\simeq 0.478$ (and eccentric ones if $\varepsilon<\varepsilon_{\rm crit}$). 
This shows that the VP95 formalism for the turbulent friction applied on the equilibrium tide in convective regions alone is sufficient to explain the observed time scales of circularisation of binary stars hosting giant-components and the corresponding eccentricity distribution. It is thus not necessary to invoke the action of the dynamical tide, which is predicted to be weak as demonstrated in \Section{sec:dynamicalTide} in the case of tidal inertial waves in RGB stars (see \Figure{fig:RTeff}). A similar result is found for the planets (Fig.\,\ref{fig:eccentricities}).  Yet, the circularisation time scales are longer by several orders of magnitude (Fig.\,\ref{fig:eccentricities}, right panel).

With KIC\,2720096 ($e$=0.49,\,$P_{\rm orb}$=26.7\,d) and KIC\,8095275 ($e$=0.32,\,$P_{\rm orb}$=23.0\,d) we find two eccentric heartbeat systems with $\varepsilon_{\rm r}$\,$\gg$\,\varC if a mass ratio 1/5\,$\leq$\,$q$\,$\leq$\,1 is assumed. A lower value of $q$ would place these systems closer or even below to $\varepsilon_{\rm crit}$.
Comparing the power spectra of a synthetic and the observed light curve (following B18), indeed revealed no signature of the potential tidal inertial waves in the low-frequency regime. 
The VP95 formalism does not forbid such objects with $\varepsilon_{\rm r}$\,$\gg$\,\varC, but because time scales for circularisation are short for such systems,   we argue that they could be in a short lived phase of rapid circularisation during the evolution of some binary systems. We refer to them as 'forbidden binaries', \hbox{in analogy to \cite{Kroupa1995}.
}

To investigate theoretically which tide dominates on the RGB, we define the quantity $\delta$\,=\,$\log(\mathcal{D}_{\rm eq}/<$$\mathcal{D}$$>$$_\omega)$ as the logarithm of the ratio between the dissipation of the equilibrium and the dynamical tide (Appendix\,\ref{AppendixB} \& \ref{AppendixA}).
Most stars with measured rotation periods originate from the G16 sample, in which the effective temperature and hence radius could be overestimated (see text above). To compensate for this complication in the estimate of the dissipation ratio, we computed the evolution of $\delta$ for each star with known $P_{\rm rot}$ along the RGB phase of our non-rotating stellar tracks (see Fig. \ref{fig:RTeff}). 
Using the $10^{th}$ percentile $\delta_{10}$ (i.e., the value below which 10\% of the distribution may be found) of each $\delta$ evolution, we found that stars in the {\it Kepler} sample with known  $P_{\rm rot}$, and regardless of the stellar mass, have 90\% of chance to have at least a $\delta$\,=\,2.3. This means that the equilibrium tide clearly dominates over the dynamical tide on the RGB and consequently also in the more advanced red-giant phases. Note that taking into account the effect of the stellar rotation on the effective turbulent friction \citep{Mathis16} significantly reduces $\mathcal{D}_{\rm eq}$ \citep[see][]{Strugareketal2017} while still producing $\delta_{10} > 0$. \Table{tab:10th} shows $\delta_{\rm 10}$ and the corresponding convective timescale $\tau_{\rm conv,10}$ (computed in the middle of the convective envelope) for the subsample of \kepler stars. In addition, most of the systems in our sample have a tidal period ($\rm P_{\rm tide}\equiv 2\pi/\omega$),
with $\omega$ the principal tidal frequency defined in Appendix\,\ref{AppendixB}, 
longer than the possible range of convective turnover timescale computed from our stellar tracks. 
In these systems ($P_{\rm tide}$\,$\geq$\,$12$-$14$\,d, 1.0$\lesssim$$M$/$M_{\odot}$$\lesssim$2.2), the reduction factor for the eddy viscosity can thus be neglected.

\vspace{-2mm}
\section{Synergies with space photometry} \label{sec:synergies}
The unparalleled quality of the photometric measurements collected by the \Kepler mission allows for a discussion of the tidal interaction in the \Kepler sample beyond the mere comparison with VP95. The detection of solar-like oscillations enables a thorough characterisation of the oscillating binary component (e.g. F13, G14, G16, B14, B18), as well as the detection of surface activity, modulated by rotation (as depicted in \Figure{fig:inertialWaves}).  
Furthermore, asteroseismology of giant stars allows one to avoid systematic uncertainties, introduced by reddening corrections and the sample is not limited by the turn-off as it is the case for observations of cluster stars. 
By comparing surface rotation rates, derived from \Kepler photometry \citep[e.g.][]{Garcia2014b,Ceillier2017} with the orbital periods, G14 associated short-period systems with mode depletion. 
When the surface rotation and orbital motion become synchronised in close binaries, the red-giant component could be spun up and the magnetic field generated by the dynamo \citep{Priviteraetal2016} can be responsible for the mode damping. Figure\,\ref{fig:eccentricities} shows that these four non-oscillating primaries are located in short period and quasi-circularised systems (15\,$\leq$\,$P_{\rm orbit}$\,$\leq$\,45\,d and $e$\,$\leq$\,0.02).

As for the orbital eccentricity, \varC serves as a proxy rather than a sharp cutoff between oscillating and non-oscillating systems, whereby the eccentricity plays an important role. 
Stars with no or suppressed oscillation amplitudes are all systems with $e$$\simeq$0 at \hbox{$\varepsilon_{\rm r}$\,$\gg$\,\varC}, as shown in the right hand panel of  \Figure{fig:eccentricities}. This suggests that for quasi-circularised systems that \varC acts also as a limit between oscillating stars and stars with additional damping or complete suppression of oscillations. The only primary of a quasi-circularised system with detected oscillations is KIC\,5308778 ($e$$\simeq$0.006), which shows very small oscillation amplitudes, compared to compatible red-giant oscillators.  With a difference of $\sim$1.6\,d between $P_{\rm orb}$ and $P_{\rm rot}$, KIC\,5308778 (see Fig.\ref{fig:inertialWaves} \Table{tab:LiteratureParameters}) is indeed not yet fully spun up at the surface, which is the first layer to be synchronised with the orbital motion \cite[e.g.][and references therein]{Zahn2013}, indicating that the deeper layers may be not synchronised as well. The forbidden binaries KIC\,2720096 ($\varepsilon_{\rm r}$\,$\simeq$\,0.83) and KIC\,8095275 ($\varepsilon_{\rm r}$\,$\simeq$\,1.86) are both oscillating. \KIC{2720096} is showing depressed dipole modes with respect to the radial modes and exhibits the highest chromospheric activity found in the sample of B14. The oscillation amplitude of KIC\,8095275 is normal. The tidal interaction affects the surface magnetism through the rotation-driven stellar dynamo. 
Since the surface magnetism is the responsible of the low amplitudes of the modes \citep[e.g. G14,][]{Garcia2014b}, the connections with tidal interactions is thus an indirect one.

From their diagnostic diagrams (Fig.\,5 in VP95), VP95 note that for primaries with radii between the maximum stellar radius at the tip of the RGB and the minimum radius in the red-clump (RC) phase, an ambiguity between evolutionary states exists. The high dependence of \varR on the stellar radius (Eq.\,\ref{eq:IofR}) suggests that H-shell burning stars, which are ascending the RGB should have been circularised when they reach the RC phase of quiescent He-core burning. 
Therefore, VP95 proposed the orbital eccentricity as a diagnostic to distinguish between the RGB and RC evolutionary stages,  suggesting that circularised systems host RC stars.  Based on the detection of mixed-dipole modes from high precision space photometry \citep{Beck2011},  asteroseismic techniques have been developed to discriminate between RGB and RC giants \citep[e.g.][]{Bedding2011, Kallinger2014}. Applying such techniques, B14 found that all heartbeat systems in their sample 
are located on the RGB, while G14 found a mix of evolutionary states among their  binaries. In a detailed reanalysis, \cite{Kallinger2018} found that all except one \textit{oscillating} stars from G14 (KIC\,9246715, $e$\,$\sim$0.36, see also R16) belong to the less advanced RGB phase. This includes the only circularised oscillating system KIC\,5308778. 

With 14\,\Rsun, KIC\,4569590 could either be a RC star or a RGB star. On the other hand, the primary's radius from light curve models (G16) of the three other systems ($\sim$8\,\Rsun) are only compatible with the stars belonging to the low-luminosity part of the RGB. 
To test if a more eccentric version of the system would have hosted a red-giant star at the tip of the RGB, we calculated tracks of constant angular momentum in the P$_{\rm orbit}$-$e$ plane \citep{Mazeh2008} for a system with a circular orbit periods, P$_{\rm circ}$,
\begin{equation}
{P}_{\rm circ} = {P}\left(1-e^2\right)^{3/2}\,.
\end{equation}
The possible area for eccentric versions of the now-circularised \Kepler stars in this diagram is shown as red surface in the 
left hand panel of \Figure{fig:eccentricities}. We note that the track for $P_{\rm circ}$\,=\,15\,d describes the outer envelope for the distribution of binaries. Therefore, a progenitor system with $e$\,$\simeq$\,0.7 would have had a period of about 100 days. Given the close separation at periastron, such system will eventually undergo a common envelope phase on the low-luminosity RGB (B14). 
Therefore,  the orbital eccentricity cannot serve as a diagnostic of the evolutionary state for stars. 

\vspace{-2mm}
\section{Conclusions}\label{sec:Conclusions}
We tested the theory of tidal interaction on binary systems that host a red-giant component and are well characterised through \Kepler photometry. We found that the action of the turbulent friction acting on the equilibrium tide in the deep convective envelope of RGB stars allows us to understand the eccentricity-period distribution. Dissipation of tidal inertial waves is thus not required to explain the observed eccentricities in the analysed sample of stars. From our calculations, based on models of \citet{remus2012} and \cite{Gallet17}, we show that this component of the dynamical tide is expected to be weak when compared to the equilibrium tide. This is confirmed by the lack of observational signatures (B18). Therefore, the dynamical tide has a negligible contribution to the tidal dissipation budget in evolved stars with large convective envelopes

This study benefits from  complementary observational constraints, provided by the \Kepler satellite. Therefore, we could extend the discussion of \varR beyond eccentricity and use the presence of stellar oscillation. Knowing the surface rotation, allows us  to test for synchronisation of the primary with orbital motion. From this approach, we suggest that the previously reported mode depletion 
is connected indirectly, via a rotation-driven dynamo, to strong tidal interactions on short circularisation time scales (\varR\,$>$\,\varC). Consequently, the control physical parameter is the volume of the convective envelope of the giant component. Using asteroseismic constraints, we showed that eccentricity cannot serve as a diagnostic for the evolutionary state of the primary.

\section*{Acknowledgements}
We thank the referee for useful comments that allowed us to improve the article. PGB acknowledges support by the MINECO-prog.\,'Juan de la Cierva Incorporacion'\,(IJCI-2015-26034). RAG  acknowledges the ANR,\,France,\,prog.\,IDEE (n$^\circ$ANR-12-BS05-0008). StM acknowledges the ERC through SPIRE (grant\,No.\,647383). RAG and StM acknowledge support from CNES PLATO grant at CEA. FG acknowledges support from the CNES fellowship. FG\,\&\,CC acknowledge support from the Swiss National Science Foundation \& SEFRI for proj.\,C.140049 under COST\,Action\,TD\,1308\,Origins. 

%\vspace{-2mm}
\bibliographystyle{mnras}
\bibliography{bibliographyTidesMNRAS.bib}

%==================================================================
\newpage
\begin{appendix}

\section{Efficiency of the equilibrium tide dissipation} \label{AppendixB}

The dissipation of the equilibrium tide can be estimated using the theoretical
model developed by \citet{remus2012}. They obtained
\begin{equation}
  \label{eq:q_eq_phys}
  {\mathcal D}_{\rm eq} = 4\pi\frac{2088}{35} \frac{R_\star^4}{G M_\star^2}
  \left| \omega \int_{\alpha}^{1} x^8 \rho \nu_t \, {\rm d}x \right|\, ,
\end{equation}
where $\omega$ is the tidal frequency, which is evaluated here using $\omega =2\left(\Omega_{\rm orb}-\Omega_{\star}\right)$ the principal tidal frequency with $\Omega_{\rm orb}=2\pi/P_{\rm orb}$ the orbital frequency of the companion and $\Omega_{\star}\,=\, 2\pi/P_{\rm rot}$ the stellar surface rotation frequency. We introduced $G$ the gravitational constant, $\nu_t$ the effective turbulent viscosity applied on tides in the convective zone, $\alpha = R_{\rm c}/R_\star$, $R_{\rm c}$ being the radius of its base and $R_\star$ the radius of the star, $\rho$ the density in the convection zone, and $x=r/R_\star$ the normalised radial coordinate. 

We derive hereafter an estimate of Eq. \eqref{eq:q_eq_phys} as a function of the global parameters of the system.

The turbulent viscosity strength depends on how the convective turnover time ${\tau}_{\rm conv}$ compares to the tidal frequency $\omega$. We here follow \citet{remus2012} and we get
\begin{equation}
  \label{eq:turb_regimes}
  \nu_t = \frac{1}{3}v_{\rm conv} l_{\rm conv} \left[1+\left(\frac{\tau_{\rm conv}\omega}{\pi}\right)^2 \right]^{-1/2}\, ,
\end{equation}
where $v_c$ is the typical convective velocity and $l_{c}$ the mixing
length. We here assume the linear attenuation of the effective turbulent friction applied on tides for short tidal period ($P_{\rm tide}=2\pi/\omega$) as proposed by \cite{zahn1966}. In the case where we neglect the action of rotation on the effective turbulent friction applied to the equilibrium tide \citep{Mathis16} because we compute here non-rotating stellar models, we can write \citep[e.g.][]{Brunetal2015}
\begin{eqnarray}
  \label{eq:Vc}
  v_{\rm conv} &\simeq& \left(\frac{L_\star}{{\rho}_{\rm c}R_\star^2}\right)^{1/3}, \\
  \label{eq:lc}
  l_{\rm conv} &\simeq& \alpha_{\rm MLT} H_P, \\
  \label{eq:tc}
  {\tau}_{\rm conv} %\left(R_o^c\right) 
  &=& \frac{l_{\rm conv}}{v_{\rm conv}} %\alpha_{\rm MLT} H_P
          %\left(\frac{L_\star}{\bar{\rho}_{CZ}R_\star^2}\right)^{-1/3}
          \, ,
\end{eqnarray}
where we have introduced the stellar luminosity $L_\star$, the average
density in the convection zone ${\rho}_{\rm c}$, the mixing-length
parameter $\alpha_{\rm MLT}$, and the pressure scale height $H_P$. To provide an order of magnitude of the dissipation as a function of stellar global parameters, we assume the simplest approximations that $\rho={\rho}_{\rm c}$ and that $\nu_t$ in the integral in Eq.~\eqref{eq:q_eq_phys} does not vary with depth. In addition, we approximate the mixing length $l_{c}$ %$\alpha_{\rm MLT}H_P$ 
by its maximum which is given by the depth of the convection zone $(1-\alpha)R_\star$. Finally, we express the mean density of the convective envelope as a function of the mass ratio $\beta=M_{\rm c}/M_{\star}$, where $M_{\rm c}$ is the radiative core mass. We get ${\rho}_{\rm c} = 3 M_\star\left(1-\beta\right)/4\pi R_\star^3\left(1-\alpha^3\right)$ that leads to
\begin{equation}
  \label{eq:dissip_end}
  {\mathcal D}_{\rm eq} \simeq \frac{232}{35} \frac{\left| {\tau}_{\rm conv}\,\omega \right|}{\sqrt{1+\left(\displaystyle{\frac{{\tau}_{\rm conv}\,\omega}{\pi}}\right)^2}} \frac{R_\star}{G M_\star}v_{\rm conv}^2 (1-\beta)\frac{1-\alpha^9}{1-\alpha^3}  \, 
  .
\end{equation}

\section{Frequency-averaged dissipation of the dynamical tide}\label{AppendixA}

In the formalism of \citet{Ogilvie2013} and \citet{Mathis2015}, the stellar convective envelope is assumed to be in solid-body rotation with a uniform angular velocity $\Omega_{\star}$. {The assumption of such flat rotation profile in the envelope is supported by observational and theoretical studies of the rotation profile in the convective envelope of RGB stars \citep[][respectively]{Beck2017b,diMauro2016}.}
{Moderate rotation is assumed, i.e.,} the squared ratio of $\Omega_{\star}$ to the critical angular velocity $\Omega_{\rm{crit}}$ is such that 
\begin{flalign}
\left(\Omega_{\star}/\Omega_{\rm{crit}} \right)^2 =  \left(\frac{\Omega_{\star}}{\sqrt[]{\mathcal{G} M_{\star}/R_{\star}^3}} \right)^2 \equiv \epsilon^2 < 1\,.
\end{flalign}
The effects of the centrifugal acceleration are thus neglected. We use two-layer models following \citet{Ogilvie2013} and \citet{Mathis2015} to evaluate the frequency-averaged tidal inertial waves dissipation in the stellar convective envelope and we focus on solar-metallicity stars with initial masses between 0.3 and 1.4 $M_{\odot}$. In this mass range, the convective envelope surrounds the radiative core of radius $R_{\rm{c}}$ and mass $M_{\rm{c}}$. In our two-layer model, both core and envelope are assumed to be homogeneous with respective average densities $\rho_{\rm{c}}$ and $\rho_{\rm{e}}$. This allows an analytical treatment of the problem. Such a two-layer model is commonly used in the representative forward-modelling approach in asteroseismology. Recent theoretical and observational results by \cite{diMauro2016} and \cite{Beck2017b}, respectively, have confirmed that this simplification can be used.

In the case of a coplanar binary system in which the orbit of the planet is circular, 
this frequency-averaged tidal dissipation is given by (see Eq. B4 from \citealt{Ogilvie2013}),
\begin{flalign}
\label{dissipequa}
<\mathcal{D}>_\omega =& \int^{+\infty}_{-\infty} \rm{Im}\left[k_2^2(\omega)\right] \frac{\ddm\omega}{\omega} = \frac{100\pi}{63}\epsilon^2 \left( \frac{\alpha^5}{1-\alpha^5} \right)  \left( 1-\gamma \right)^2 \\ 
\times&  \left( 1-\alpha \right)^4  \left( 1 +2\alpha+3\alpha^3 + \frac{3}{2}\alpha^3 \right)^2 \left[ 1+  \left( \frac{1-\gamma}{\gamma} \right) \alpha^3  \right] \nonumber \\ 
\times& \left[   1 + \frac{3}{2}\gamma + \frac{5}{2\gamma}  \left( 1 + \frac{1}{2}\gamma - \frac{3}{2}\gamma^2 \right)  \alpha^3 - \frac{9}{4} \left( 1-\gamma \right)\alpha^5   \right]^{-2} \nonumber,
\end{flalign}
with 
\begin{flalign}
{\rm~} \gamma = \frac{\rho_{\rm{e}}}{\rho_{\rm{c}}}=\frac{\alpha^3(1-\beta)}{\beta(1-\alpha^3)} < 1.
\end{flalign}
As in \cite{Mathis2015}, we can express a structural frequency-averaged dissipation: 
\begin{eqnarray}
\label{dissipequabis}
< \mathcal{D}>_{\rm struct.} =  \epsilon^{-2} < \mathcal{D}>  =    \epsilon^{-2} < \rm{Im}\left[k^2_2(\omega)\right] >_{\omega},
\end{eqnarray}
that only depends on $\alpha$ and $\beta$. 

We know that tidal inertial waves dissipation can vary over several orders of magnitude as a function of the forcing frequency \citep[][]{OgilvieLin2007}. As discussed in \cite{Mathis2015}, the frequency-averaged dissipation thus provides us a qualitative order of magnitude of this dissipation for a rotating star at a given evolutionary stage.

We note that here we use prescriptions for the dynamical tide derived for a coplanar and circular system. In this simplified case, the tidal frequency is the principal one given by $\omega =2\left(\Omega_{\rm orb}-\Omega_{\star}\right)$. Moreover, in that configuration the tidal potential is described by the spherical harmonic of degree l=2, m=$\pm$2 and thus by the love number $k_2^2$. Taking into account coherently the eccentricity will require to consider general tidal frequencies $\omega=L\Omega_{\rm orb}-m\Omega_{\star}$ (where $L$ is an integer number) that include the principal tidal frequency for which $L=2$ and $m=2$, and other spherical harmonic components of the tidal potential and corresponding complex Love numbers $\left(k_2^0\right)$ \citep{Zahn1977,Ogilvie2014}.

\section{Literature values for \Kepler-systems}\label{sec:LiteratureValues}

The parameters for eclipsing binaries are taken from the dynamical solution of G16. For the list of heartbeat stars (B14), radius and mass were inferred from seismic scaling relations and corrected for the systematic mass overestimate of 15\% reported by G16. Surface rotation periods were adopted from G14, B14 and B18. For four stars of B14, KIC\,7431665, KIC\,11044668, KIC\,8803882, and KIC\,7799540,  no orbital parameters have yet been published.

\subsection{Content of \Table{tab:LiteratureParameters}}

We refer to the cited literature for details on the applied methodology.
\Table{tab:LiteratureParameters} contains the following parameters,
\begin{itemize}
\item \textit{KIC} specifies the target identification number in the \Kepler Input Catalog.
\item  \textit{Type} indicates if a binary system is an eclipsing binary (EB), a heartbeat system (HB), or an eclipsing heartbeat system (eHB).	'NO' indicates that the star belongs to the four non-oscillating stars of G14/G16.
\item  P$_{\rm orbit}$ is the measured orbital period.	
\item  $e$ is the orbital eccentricity.
\item  \num is the peak frequency of the excess of oscillation power.
\item  $R$/$R_{\odot}$ is the stellar radius in solar units. Values from B14 are seismically inferred and corrected for the 5\% overestimate of seismic radius. Values from G16 originate from a dynamical solution.
\item  $M$/$M_{\odot}$ is the stellar mass in solar units. Values from B14 are seismically inferred and corrected for the 15\% overestimate of  seismic mass. Values from G16 originate from a dynamical solution.		
\item  \new{$q=M_2$/$M_1$} is the mass ratio between the two stellar components in the system. '?' indicates if $q$ has not been determined for a given system. 
\item  $T_{\rm eff}$ is the effective temperature. 
\item  \textit{P$_{\rm rot}$} specifies the time scale of the flux modulation, identified as the surface rotation period. The sources of the values are G14 and B18. We round all period values to the next full day.
\item  The dimensionless number \textit{\varR} is proportional to the inverse of the time scale of tidal circularisation (see Eq. \ref{eq:varr}). If no value of the mass ratio is specified, \varR is calculated for $q=0.5$.
\item REF: the last column is specifying the literature references. If several papers are reporting on a given system, values of the most recent paper are cited. Previous references are given in brackets.
\end{itemize}

\subsection{Content of \Table{tab:10th}}

\Table{tab:10th} lists the following parameters for the systems' red-giant primary with known $P_{\rm rot}$,
\begin{itemize}
\item $\delta_{\rm 10}$ is the 10$^{\rm th}$ percentile of the logarithm of the ratio between the dissipation of the equilibrium and the dynamical tide, $\delta$\,=\,$\log (\mathcal{D}_{\rm eq}$\,/\,$<$$\mathcal{D}$$>$$_\omega)$.
\item $\tau_{\rm conv,10}$ is the corresponding convective turnover timescale computed in the middle of the convective zone.
\item $P_{\rm tide}$ > $\tau_{\rm conv,10}$ indicates if the tidal period is longer than the convective turnover timescale  (Yes / No).

\end{itemize}

\newpage
\setcounter{table}{1}
\begin{table}
\caption{Evolution of $\delta_{\rm 10}$ for the systems with know rotation period.}
\label{tab:10th}
\centering
\tabcolsep=5pt

\begin{tabular}{rrccc}
\hline\hline
\multicolumn{1}{c}{KIC} & 
\multicolumn{1}{c}{$P_{\rm tide}$} & $\tau_{\rm conv,10}$ & $P_{\rm tide}$ > $\tau_{\rm conv,10}$ & $\delta_{\rm 10}$\\
 & \multicolumn{1}{c}{[days]} & [days] & & \\
\hline
7943602 & 357 & 20 & Y & 2.3-3.4  \\
7377422 & 56 & 20 & Y & 3.7-4.8\\
3955867 & 845 & 20-23 & Y & 2.6-3.7\\
9291629 & 692 & 20-23 & Y & 2.3-3.4 \\
5179609 & 28 & 23 & Y & 4.7-5.8 \\
9163796 & 906 & 23 & Y & 3.8-4.9 \\
8430105 & 65 & 23-26 & Y & 4.4-5.5 \\
5308778 & 505 & 12 & Y & 3.0-4.1\\
4569590 & 2285 & 12 & Y & 2.4-3.5\\
8702921 & 12 & 12 & N & 4.2-5.3\\
9246715 & 101 & 14 & Y & 4.1-5.2 \\
\hline
\end{tabular}
\end{table}

\setcounter{table}{0}

\begin{landscape}
\begin{table}
\caption{Literature Parameters of red-giant binaries in the \Kepler sample. }
\label{tab:LiteratureParameters}
\centering
\tabcolsep=5pt
\begin{tabular}{rrrrrrrrrrrl}
\hline\hline								
\multicolumn{1}{c}{KIC}	&	
\multicolumn{1}{c}{Type}	&	
\multicolumn{1}{c}{P$_{\rm orb}$}	&	
\multicolumn{1}{c}{$e$}	&	
\multicolumn{1}{c}{\num}		&	
\multicolumn{1}{c}{R/R$_\odot$}	&	
\multicolumn{1}{c}{M/M$_\odot$}	&	
\multicolumn{1}{c}{$q$}	&	
\multicolumn{1}{c}{T}	&	
\multicolumn{1}{c}{P$_{\rm rot}$}	&	
\multicolumn{1}{c}{\varR}	&	
REF	\\
	&		&	
\multicolumn{1}{c}{[days]}	&	
\multicolumn{1}{c}{[]}	&
\multicolumn{1}{c}{[$\mu$Hz]}	&	
\multicolumn{1}{c}{[]}	&	
\multicolumn{1}{c}{[]}	&	
\multicolumn{1}{c}{[]}&	
\multicolumn{1}{c}{[K]}	&	
\multicolumn{1}{c}{[days]}	&
\multicolumn{1}{c}{[]}	&		\\
\hline
	2444348	&	HB	&	103.50	$\pm$	0.01	&	0.48	$\pm$	0.01	&	30.5	$\pm$	0.3	&		14.2	$\pm$	0.3	&		1.6	$\pm$	0.1	&	?			&	4565	&	-	&	-0.53	&	B14	\\	%	14.9	1.94			-0.29264836	23617660210411000.0	0.5										
	2697935	&	eHB	&	21.50	$\pm$	0.02	&	0.41	$\pm$	0.02	&	$\sim$	405.6		&	$\sim$	3.1			&	$\sim$	1.2			&	?			&	4883	&	-	&	-0.73	&	B14	\\	%	3.26	1.45			-0.187737261	1193550610212.7	0.5										
	2720096	&	HB	&	26.70	$\pm$	0.01	&	0.49	$\pm$	0.01	&	110.1	$\pm$	0.7	&		6.6	$\pm$	0.1	&		1.3	$\pm$	0.1	&	?			&	4812	&	-	&	0.83	&	B14	\\	%	6.98	1.54			-6.735778437	169548153353192.0	0.5										
	3955867	&	EB, NO	&	33.65685	$\pm$	0.00007	&	0.019	$\pm$	0.002	&	$-$			&		7.9	$\pm$	0.1	&		1.10	$\pm$	0.06	&	0.84	$\pm$	0.05	&	4884	&	33	&	1.14	&	G16 (G14)	\\	%			0.92	0.03	-13.83929209	530109950860996.0	0.836363636										
	4569590	&	EB, NO	&	41.3710	$\pm$	0.0001	&	0.004	$\pm$	0.001	&	$-$			&		14.1	$\pm$	0.2	&		1.6	$\pm$	0.10	&	0.66	$\pm$	0.05	&	4706	&	41	&	1.67	&	G16 (G14)	\\	%			1.05	0.04	-47.17584985	23026610543707600.0	0.65625										
	4663623	&	EB	&	358.09	$\pm$	0.0003	&	0.43	$\pm$	0.01	&	54.1	$\pm$	0.2	&		9.7	$\pm$	0.2	&		1.36	$\pm$	0.09	&	0.99	$\pm$	0.08	&	4812	&	-	&	-4.08	&	G16 (G14)	\\	%			1.34	0.07	-8.34456E-05	2016961378909580.0	0.985294118										
	5006817	&	HB	&	94.812	$\pm$	0.002	&	0.71	$\pm$	0.01	&	145.9	$\pm$	0.5	&		5.5	$\pm$	0.1	&		1.3	$\pm$	0.1	&	0.199	$\pm$	0.001	&	5000	&	-	&	-2.80	&	B14	\\	%	5.84	1.49			-0.001598043	53105792890595.8	0.199										
	5039392	&	HB	&	236.70	$\pm$	0.02	&	0.44	$\pm$	0.01	&	6.2	$\pm$	0.1	&		22.8	$\pm$	0.7	&		0.8	$\pm$	0.1	&	?			&	4110	&	-	&	-0.01	&	B14	\\	%	24	0.98			-0.967120434	525980261009751000.0	0.5										
	5179609	&	EB	&	43.93108	$\pm$	0.000002	&	0.150	$\pm$	0.001	&	322	$\pm$	1.0	&		3.50	$\pm$	0.03	&		1.18	$\pm$	0.03	&	0.51	$\pm$	0.02	&	5003	&	182	&	-1.96	&	G16 (G14)	\\	%			0.60	0.01	-0.01088514	2646650293458.1	0.508474576										
	5308778	&	EB	&	40.5661	$\pm$	0.0003	&	0.006	$\pm$	0.005	&	49	$\pm$	1.1	&		10.1	$\pm$	0.3	&		1.5	$\pm$	0.1	&	0.43	$\pm$	0.03	&	4900	&	39	&	0.80	&	G16 (G14)	\\	%			0.64	0.01	-6.305804877	2623890671999020.0	0.426666667										
	5786154	&	EB	&	197.918	$\pm$	0.0004	&	0.3764	$\pm$	0.0009	&	29.8	$\pm$	0.2	&		11.4	$\pm$	0.2	&		1.06	$\pm$	0.06	&	0.96	$\pm$	0.07	&	4747	&	-	&	-1.85	&	G16 (G14)	\\	%			1.02	0.04	-0.014007755	5771174080581860.0	0.962264151										
	7037405	&	EB	&	207.1083	$\pm$	0.0007	&	0.238	$\pm$	0.004	&	21.8	$\pm$	0.1	&		14.1	$\pm$	0.2	&		1.25	$\pm$	0.04	&	0.91	$\pm$	0.03	&	4516	&	-	&	-1.62	&	G16 (G14)	\\	%			1.14	0.02	-0.023721231	23026610543707600.0	0.912										
	7377422	&	EB	&	107.6213	$\pm$	0.0004	&	0.4377	$\pm$	0.0005	&	40	$\pm$	2.1	&		9.5	$\pm$	0.2	&		1.05	$\pm$	0.08	&	0.81	$\pm$	0.07	&	4938	&	55	&	-0.96	&	G16 (G14)	\\	%			0.85	0.03	-0.109828309	1761141547380960.0	0.80952381										
%	7431665	&	HB	&	281.4	$\pm$	?	&	-	$\pm$	-	&	54.0	$\pm$	0.7	&		8.9	$\pm$	0.1	&		1.2	$\pm$	0.1	&	?			&	4580	&	-	&	-3.59	&	B14	\\	%	9.4	1.36			-0.000258751	1177217175096960.0	0.5										
%	7799540	&	HB	&	71.8	$\pm$	?	&	-	$\pm$	-	&	347.2	$\pm$	5	&	$\sim$	3.5			&	$\sim$	1.3			&	?			&	5177	&	-	&	-3.28	&	B14	\\	%	3.64	1.52			-0.000521441	2446607317192.6	0.5										
	7943602	&	EB, NO	&	14.69199	$\pm$	0.00004	&	0.001	$\pm$	0.003	&	$-$			&		6.6	$\pm$	0.2	&		1.0	$\pm$	0.10	&	0.78	$\pm$	0.09	&	5096	&	15	&	2.70	&	G16 (G14)	\\	%			0.78	0.05	-497.4981512	164454064905904.0	0.78										
	8054233	&	EB	&	1058.16	$\pm$	0.02	&	0.2718	$\pm$	0.0004	&	46.5	$\pm$	0.3	&		10.7	$\pm$	0.1	&		1.60	$\pm$	0.06	&	0.69	$\pm$	0.04	&	4971	&	-	&	-6.61	&	G16 (G14)	\\	%			1.10	0.04	-2.46436E-07	3820339730270180.0	0.6875										
	8095275	&	HB	&	23.00	$\pm$	0.01	&	0.32	$\pm$	0.01	&	69.3	$\pm$	0.3	&		7.4	$\pm$	0.1	&		1.0	$\pm$	0.1	&	?			&	4622	&	-	&	1.86	&	B14	\\	%	7.78	1.21			-73.23309501	343614465988859.0	0.5										
	8144355	&	HB	&	80.55	$\pm$	0.01	&	0.76	$\pm$	0.01	&	179.0	$\pm$	2	&		4.7	$\pm$	0.1	&		1.1	$\pm$	0.1	&	?			&	4875	&	-	&	-2.41	&	B14	\\	%	4.9	1.26			-0.003890391	16942236650096.4	0.5										
	8210370	&	HB	&	153.50	$\pm$	0.01	&	0.70	$\pm$	0.01	&	44.1	$\pm$	0.8	&		10.0	$\pm$	0.2	&		1.2	$\pm$	0.1	&	?			&	4585	&	-	&	-1.92	&	B14	\\	%	10.5	1.4			-0.012117691	2419561133683660.0	0.5										
	8410637	&	EB	&	408.3	$\pm$	0.5	&	0.689	$\pm$	0.001	&	46.0	$\pm$	0.2	&		10.7	$\pm$	0.1	&		1.56	$\pm$	0.3	&	0.85	$\pm$	0.16	&	4800	&	-	&	-4.34	&	F13	\\	%			1.32	0.02	-4.60123E-05	3820339730270180.0	0.846153846										
	8430105	&	EB	&	63.32713	$\pm$	0.00003	&	0.2564	$\pm$	0.0002	&	76.7	$\pm$	0.6	&		7.65	$\pm$	0.05	&		1.31	$\pm$	0.02	&	0.63	$\pm$	0.01	&	5042	&	122	&	-0.73	&	G16 (G14)	\\	%			0.83	0.01	-0.187065661	429981267052472.0	0.633587786										
	8702921	&	EB	&	19.38446	$\pm$	0.00002	&	0.0964	$\pm$	0.0008	&	195.6	$\pm$	0.5	&		5.32	$\pm$	0.05	&		1.67	$\pm$	0.05	&	0.16	$\pm$	0.01	&	5058	&	98	&	0.26	&	G16 (G14)	\\	%			0.274	0.009	-1.815382661	40410880990330.7	0.164071856										
%	8803882	&	HB	&	89.7	$\pm$	?	&	-	$\pm$	-	&	347.0	$\pm$	3	&		3.5	$\pm$	0.1	&		1.2	$\pm$	0.1	&	?			&	5043	&	-	&	-3.64	&	B14	\\	%	3.68	1.4			-0.00023094	2627021107627.9	0.5										
	8912308	&	HB	&	20.17	$\pm$	0.01	&	0.23	$\pm$	0.01	&	$\sim$	350.0	0	&	$\sim$	4.0			&	$\sim$	1.7			&	?			&	4872	&	-	&	-0.39	&	B14	\\	%	4.2	2.02			-0.407215935	6210794263470.3	0.5										
	9151763	&	HB	&	437.51	$\pm$	0.03	&	0.73	$\pm$	0.01	&	13.8	$\pm$	0.2	&		16.7	$\pm$	0.4	&		1.0	$\pm$	0.1	&	?			&	4290	&	-	&	-2.62	&	B14	\\	%	17.6	1.19			-0.002379785	69836381728503000.0	0.5										
	9163796	&	HB	&	121.30	$\pm$	0.01	&	0.69	$\pm$	0.002	&	165.3	$\pm$	1.3	&		5.1	$\pm$	0.1	&		1.2	$\pm$	0.1	&	0.985	$\pm$	0.005	&	4960	&	130	&	-3.17	&	B18 (B14)	\\	%	5.35	1.39			-0.00066911	30017697979301.3	0.985221675										
	9246715	&	EB	&	171.27688	$\pm$	0.00001	&	0.3559	$\pm$	0.0003	&	106.4	$\pm$	0.8	&		8.30	$\pm$	0.04	&		2.149	$\pm$	0.007	&	0.990	$\pm$	0.005	&	5030	&	93	&	-3.54	&	R16 (G14)	\\	%			2.171	0.007	-0.000289876	731159241518636.0	1.01023732										
	9291629	&	EB, NO	&	20.68643	$\pm$	0.00004	&	0.007	$\pm$	0.002	&	$-$			&		7.99	$\pm$	0.05	&		1.14	$\pm$	0.03	&	0.96	$\pm$	0.03	&	4713	&	21	&	2.26	&	G16 (G14)	\\	%			1.1	0.02	-180.6275087	570680621599525.0	0.964912281										
	9408183	&	HB	&	49.70	$\pm$	0.01	&	0.42	$\pm$	0.01	&	164.8	$\pm$	0.2	&		4.8	$\pm$	0.1	&		1.0	$\pm$	0.1	&	?			&	4900	&	-	&	-1.18	&	B14	\\	%	5.02	1.23			-0.065346161	19832408593602.6	0.5										
	9540226	&	eHB	&	175.4439	$\pm$	0.0006	&	0.3880	$\pm$	0.0002	&	27.1	$\pm$	0.2	&		12.8	$\pm$	0.1	&		1.33	$\pm$	0.05	&	0.74	$\pm$	0.04	&	4692	&	-	&	-1.64	&	G16 (B14, G14)	\\	%			0.98	0.03	-0.023126672	12267342609935500.0	0.736842105										
	9970396	&	EB	&	235.2985	$\pm$	0.0002	&	0.194	$\pm$	0.007	&	63.7	$\pm$	0.2	&		8.0	$\pm$	0.2	&		1.14	$\pm$	0.03	&	0.89	$\pm$	0.03	&	4916	&	-	&	-3.38	&	G16 (G14)	\\	%			1.02	0.02	-0.000418992	575346410047194.0	0.894736842										
	10001167	&	EB	&	120.3903	$\pm$	0.0005	&	0.159	$\pm$	0.003	&	19.9	$\pm$	0.1	&		12.7	$\pm$	0.3	&		0.81	$\pm$	0.05	&	0.98	$\pm$	0.07	&	4700	&	-	&	0.03	&	G16 (G14)	\\	%			0.79	0.03	-1.077727137	11656705266394900.0	0.975308642										
	10614012	&	eHB	&	132.13	$\pm$	0.01	&	0.71	$\pm$	0.01	&	70.2	$\pm$	0.9	&		8.2	$\pm$	0.2	&		1.3	$\pm$	0.1	&	?			&	4715	&	-	&	-2.23	&	B14	\\	%	8.6	1.49			-0.005849059	659749667940527.0	0.5										
%	11044668	&	HB	&	139.5	$\pm$	?	&	-	$\pm$	-	&	50.2	$\pm$	0.2	&		7.8	$\pm$	0.1	&		0.8	$\pm$	0.1	&	?			&	4565	&	-	&	-1.85	&	B14	\\	%	8.18	0.99			-0.014152208	476231666115527.0	0.5																														
						
						\hline
\end{tabular}
\end{table}%
\end{landscape}

\end{appendix}

% Don't change these lines
\bsp	% typesetting comment
\label{lastpage}
\end{document}